\documentclass{aa}  
\usepackage{graphicx}
\usepackage{txfonts}
\usepackage{float}
\usepackage[normalem]{ulem}
\usepackage{xcolor}
\usepackage{times,lipsum}
\usepackage[colorlinks=true,linkcolor=blue,citecolor=blue, urlcolor=blue]{hyperref}

\begin{document}

   \title{Morphological variations of solar granules in the presence of magnetic fields}

   \author{J. I. Campos Rozo
          \inst{1,2}
          \and
          J. Jur\v{c}\'ak\inst{1}
          \and
          S. M. D\'{i}az Castillo\inst{3}
          \and
          M. van Noort\inst{4}
          }

   \institute{Astronomical Institute of the Czech Academy of Sciences, Ond\v{r}ejov, Czech Republic\\
              \email{jose.ivan.campos.rozo@asu.cas.cz}\\
              \email{jan.jurcak@asu.cas.cz}
        \and
             Universidad Nacional de Colombia, Observatorio Astronómico Nacional de Colombia, Bogotá, Colombia
        \and
            Institut für Sonnenphysik (KIS), Freiburg, Germany
        \and
            Max-Plank Institute for Solar System Research, G\"{o}ttingen, Germany
        }

   \date{Received XXXXX; accepted XXXX}
 
  \abstract
   {Solar granulation consists of dynamic convective plasma cells that rise from the solar interior to the surface. The interaction between this plasma cells and the Sun’s magnetic field provides valuable insights into the dynamics of plasma near the solar surface and how it changes in the presence of magnetic field.}
   {This study aims to analyse the morphological characteristics of solar convective cells, investigating the relationship between magnetic field properties and granule dynamics. In particular, we examine how granule properties, such as area, shape, and brightness, vary under different magnetic field conditions.}
   {The research used observations of the active region NOAA 11768 taken by the Swedish 1-m Solar Telescope (SST). We applied segmentation algorithm on the continuum intensity images to identify individual granules and determine their sizes, shapes, and mean brightness. We determined the magnetic field vector and line-of-sight velocity from the CRISP spectropolarimetric data to investigate the role of these parameters on the properties of granules.}
   {We found that granular area decreases systematically with increasing magnetic field strength, with the largest granules found in non-magnetic regions and a mean granule area of approximately 1.58 arcsec$^2$ with an effective diameter of 1.42 arcseconds. Both mean continuum intensity and granule size decrease with stronger magnetic fields, demonstrating the suppression of convective energy transport in magnetised regions. However, we do not find any correlation between the mean brightness of granules and mean up-flow velocity within the granules. We observe highly elongated granules in both magnetic and non-magnetic regions, but close to circular granules are observed only in non-magnetic areas. We find indications of alignment between major axis of granules and magnetic field azimuth in regions with a strong horizontal component of the magnetic field. These findings confirm that granules are highly sensitive to the presence of magnetic field, with strong fields inhibiting lateral expansion of convective cells.
   }
   {}

   \maketitle
%
\nolinenumbers
\section{Introduction}

Solar granulation is a manifestation of the subsurface convection in the solar photosphere, which presents itself as a pattern of bright cellular elements surrounded by darker intergranular lanes. These structures are important to understand the solar surface dynamics, and they have been extensively studied through both observational and numerical approaches \citep[i.e.][]{Roudier1986, Muller1989, Hirzberger1997, Danilovic2008}. The larger part of the solar surface exhibits this granulation pattern, where convective cells heat the photosphere with the hot material rising from the upper layer of the convective zone. The emerged small-scale magnetic field is dragged by such convective cells \citep{Cheung2014, Guglielmino2020}. While no singular characteristic size exists for granular cells, studies consistently report mean diameter between 1" and 2" (approximately 725-1450 km) in quiet Sun regions \citep{Woehl1985}. 

\begin{figure*}[htb]
\centering
   \includegraphics[width=18.3cm]{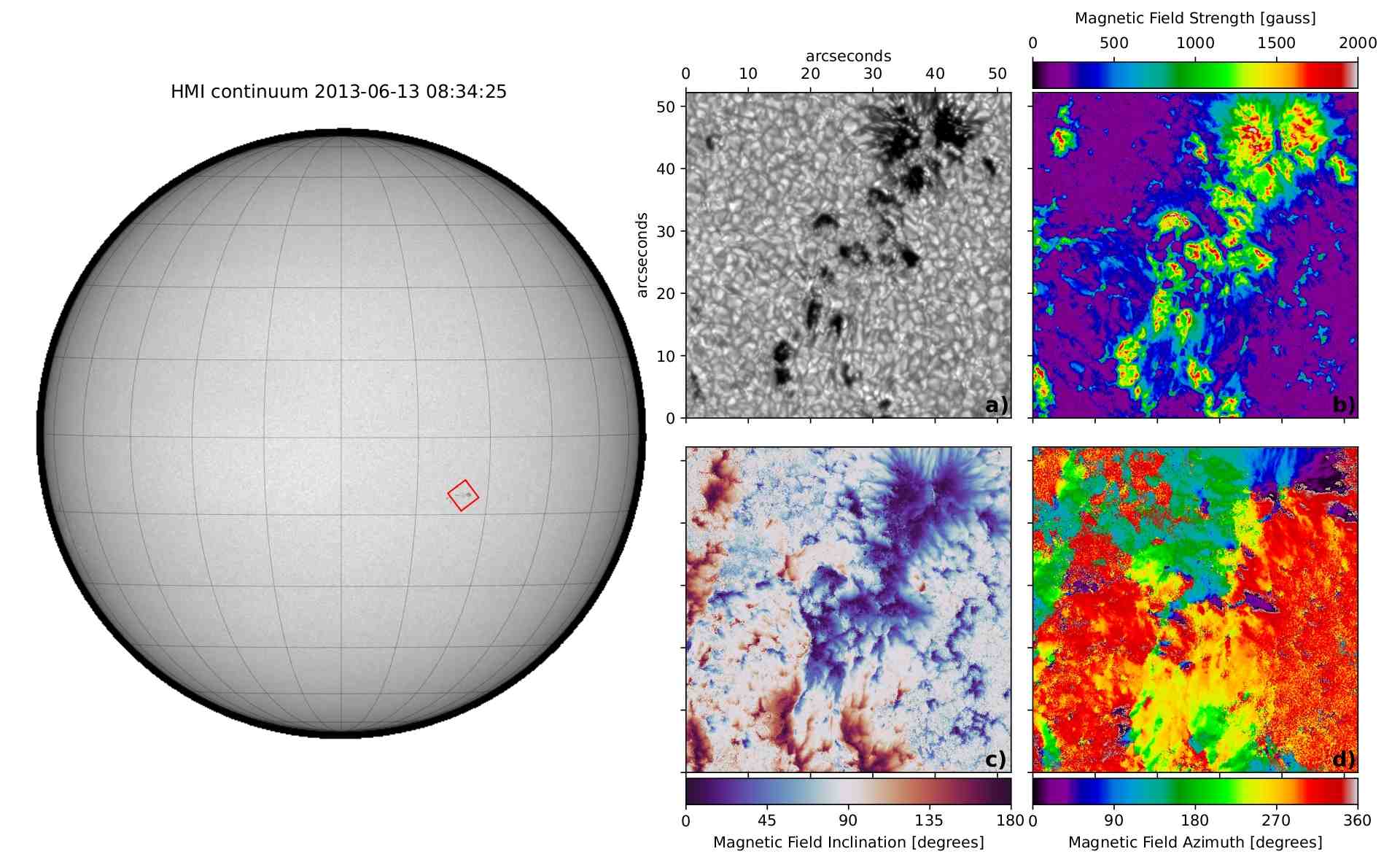}
      \caption{Left: Full-disk image context from Helioseismic and Magnetic Imager (HMI) on board of Solar Dynamics Observatory (SDO). The red rectangle shows the region of interest observed by the 1-m SST. Right: Four images grid displaying maps obtained from the SST data; a) blue-continuum intensity image, b) strength of the magnetic field, c) magnetic field inclination and d) magnetic field azimuth.
              }
         \label{context_image}
\end{figure*}

Earlier studies reported varying characteristic sizes, from 1.1" \citep{Namba1969} to 1.35" \citep{Bray1984}, and mean intergranular distances of 1.76" \citep{Roudier1986}. These granules may in turn grow and fragment, or merge with others, or shrink and decompose \citep{Bahng1961, Title86, Mehltretter1978}. However, more recent investigations have revealed a distinct population of smaller convective cells with spatial scales below 600 km \citep[e.g.][]{Roudier2003, Yu2011}, as identified both observationally and in high-resolution solar radiation hydrodynamics simulations \citep{Stein2009, Abramenko2012, Rempel2014}. 

The properties of granulation in regions of strong magnetic field exhibit additional complexity due to the interaction between such field and the convective plasma. From a quantitative point of view, studies of such interaction, like those performed by \citet{Aparna2025}, demonstrate that stronger vertical fields increasingly suppress horizontal surface motions, suggesting morphological variation of the granulation in regions of magnetic flux emergence. In the case of sunspot umbrae, magneto-convection manifests as umbral dots \citep{Sobotka2009}, while in penumbrae, where magnetic fields are weaker and more inclined, convective cells become highly elongated, forming characteristic filaments \citep{Rempel2011, Tiwari2013}. Moreover, granules in plage regions typically exhibit reduced sizes and lower line-of-sight velocities compared to field-free granulation \citep{Narayan2010}. Also, broad light bridges in sunspots show a granular pattern where the properties of convective cells are comparable to the quiet Sun granules \citep{Lagg:2014}, but they are smaller and have longer lifetimes \citep{Hirzberger:2002}. Early observations indicate that the sizes of granules are smaller than average near sunspots \citep{Schroter1964,Macris1979}, while granules can be highly elongated in regions of flux emergence \citep{Schlichenmaier:2010, Centeno2007}. Previous studies have indicated a characteristic scale diameter of approximately 1.37" that distinguishes different granular behaviours, suggesting a transition between turbulent eddies and convective elements \citep{Roudier1986}. 

The objective characterization of the fine-structure in solar granulation requires robust pattern recognition algorithms. However, the absence of a priori criteria for physically meaningful structures has led to various methodological approaches, each producing systematically different patterns.  

Traditional techniques employ Fourier-based recognition with single-level intensity thresholds. This approach has inherent limitations in structure separation \citep[i.e.][]{Roudier1986,Title1989, Hirzberger1997, Berrilli2005, bovelet2007, Yu2011}. More sophisticated methods applying machine learning techniques have been developed to address these constraints \citep[e.g.][]{Feng2013, Chola2022, Saida2022}, yet challenges remain in balancing pattern recognition accuracy with physical relevance.

The interaction between the magnetic field and plasma convective cells is a crucial aspect of solar surface dynamics, in particular, at granular scales. Using high-resolution observations, multiple studies have shown that horizontal motions inside granules carry vertical magnetic flux toward the intergranular lanes \citep{Harvey2007, Centeno2007}. The emergence of the magnetic field can disturb the granulation patterns and shapes, leading to the appearance of dark lanes or so-called abnormal granulation \citep{Cheung2007}. The statistical approach to understanding granulation patterns and their relationship with magnetic fields remains relatively unexplored. This presents an opportunity for a systematic investigation using high-resolution observational data from modern large solar telescopes and new techniques. Note that rudimentary analysis of the data was already performed by \citet{Jurcak2017}, but the presented analysis is more thorough and implements more advanced procedures for segmentation and characterization of the identified convective cells. 

\section{Observations and data processing}
   
The analysis of the active region NOAA 11768 was done using observations obtained from the 1-m Swedish Solar Telescope \citep[SST;][]{Scharmer2003}. The instruments on the SST provide full spectropolarimetric data for studying solar magnetic features at high spatial resolution as well as continuum intensity images. We analysed data from June 13, 2013, capturing a rudimentary sunspot, a number of pores and flux emergence regions within the field of view (FOV). The active region emerged on the solar surface on June 11, 2013, and was still growing during the observation period, see Fig.~\ref{context_image}.
   
The observational data comprises two main datasets. First, the blue-continuum images were processed using the Multi-Object Multi-Frame Blind Deconvolution (MOMFBD) technique \citep{vannoort2005}, achieving a temporal cadence of 5.6 s and a spatial sampling  of 0.034 arcsec pixel$^{-1}$ (see Fig.~\ref{context_image}.a). These observations were recorded between 8:21 UT and 10:50 UT. Second, the CRisp Imaging SpectroPolarimeter (CRISP) instrument \citep{Scharmer2008} collected full Stokes vector profiles of the \ion{Fe}{i}\,525\,nm line with a temporal cadence of 31 s and a spatial sampling of 0.058 arcsec pixel$^{-1}$ \citep[see measurments by][]{Noren2013} during the period between 8:36~UT and 10:54~UT. The imaging and CRISP data were carefully co-aligned to ensure precise spatial correspondence.

\begin{figure}[!t]
\centering
   \includegraphics[width=9cm]{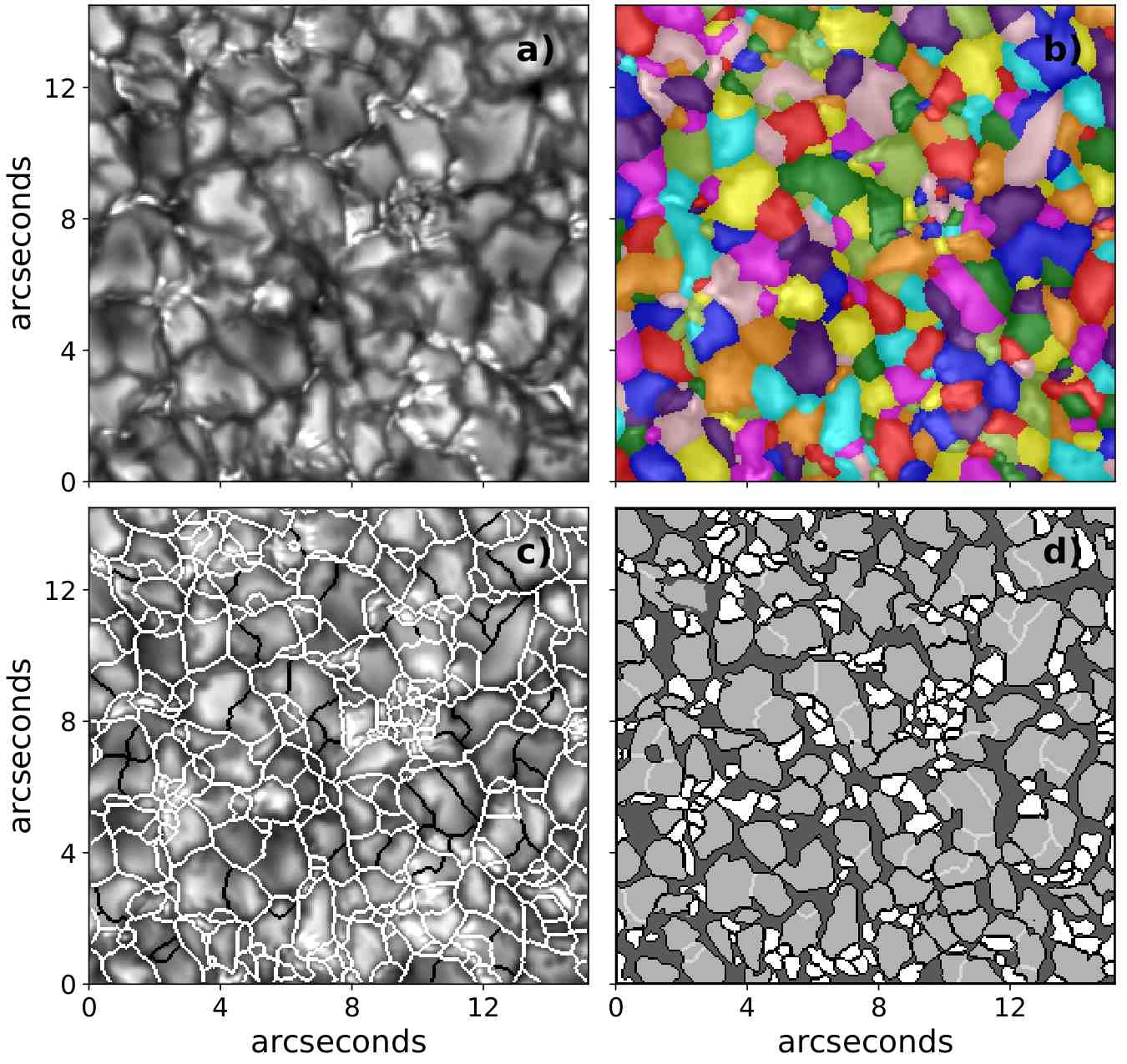}
      \caption{Segmentation process of solar photospheric granules. a) Original solar surface region. b) Initial granule segmentation using local minima and watershed technique. c) Refined contour selection after parameter optimization. d) Final granule morphology after contour-based erosion.}
         \label{segment_steps}
\end{figure}

\subsection{Inversion of the CRISP data}
\label{inversions}

To determine the magnetic field properties, we used the Very Fast Inversion of the Stokes Vector (VFISV) code by \citet{Borrero2011} to perform Milne-Eddington inversions of the Stokes vector profiles, obtaining magnetic field strength (Fig.~\ref{context_image}.b), inclination (Fig.~\ref{context_image}.c), and azimuth (Fig.~\ref{context_image}.d). The 180$^{\circ}$ ambiguity in the magnetic field azimuth problem was solved using the AMBIG code \citep{Leka2009}, followed by the transformation of the magnetic field inclination and azimuth from the line-of-sight to the local-reference frame using the AZAM routines written by Paul Seagraves at High Altitude Observatory \citep[described in][]{Lites:1995}. 

We also used the retrieved line-of-sight (LOS) velocity values in our analysis (not shown in Fig.~\ref{context_image}). The LOS velocities were corrected for convective blueshift in two steps. First, we computed the average LOS velocity in quiet-Sun regions and subtracted this value from the entire velocity map. This procedure removes velocity contributions arising from the Earth’s orbital motion and solar rotation, yielding a zero mean LOS velocity in the quiet Sun. Second, we applied a correction for convective blueshift based on the measurements and calculations of \citet{Lohner:2019}, who quantified the convective blueshift of the \ion{Fe}{i}\,525\,nm line. Taking into account the spectral resolution of CRISP and the $\mu$ angle of the observed region, we applied a uniform velocity offset of -0.33~km\,s$^{-1}$. Although this correction is tentative, it yields reasonable mean up-flow velocities in the segmented granules.

In the analysed dataset, the signal-to-noise level was around 200 for the Stokes profiles of polarised light. An intrinsic limitation of inversion codes is their tendency to fit noise. Therefore, even in the quiet areas of the FOV, we obtain a non-zero magnetic field vector. Based on the inversion results in the most quiet regions of the FOV, the magnetic field strength has a mean value of 75~G with standard deviation of 26~G. As the Stokes $Q$ and $U$ profiles are proportional to $B^2$ and Stokes $V$ is proportional to $B$, we obtain magnetic field inclination in the line-of-sight frame around 90$^\circ$ in the quiet areas where signals are dominated by noise.

\subsection{Segmentation}

The data preparation involved scaling and co-aligning the blue-continuum observations with respect to the CRISP dataset to ensure accurate spatial correspondence. Therefore, the spatial sampling of the blue-continuum images was downscaled to match the spatial sampling of CRISP data. Since our work is focused on the regions of quiet-sun and emergence of magnetic field, we masked the active region in the analysed region of interest. The initial threshold was set to 50\% of the intensity $I/I_\mathrm{QS}$, and the subsequent performance of the segmentation algorithm improves the masking of remaining penumbral regions, as these features show much lower intensity values compared to solar granulation. 

Our granulation segmentation algorithm involves a multi-step image processing to achieve an accurate identification of granules, similar to the Multi-Level Thresholding (MLT4) algorithm proposed by \citet{bovelet2007}. In the initial step (Fig.~\ref{segment_steps}.a), a small region of $250\times250$ pi\-xels ($\sim14\times14$ arcse\-conds) is extracted from the processed blue continuum intensity images. Subsequently (Fig.~\ref{segment_steps}.b) local minima detection and watershed segmentation are applied to generate a segmented map with the preliminary granule identification. By leveraging most of the MLT4 features, we refined the segmentation process by strategically modifying the segmentation parameters. The algorithm enables selective contour preservation, where white line segments are retained while black line segments are eliminated (see Fig.~\ref{segment_steps}.c). This selective filtering enhances the initial segmentation accuracy.

The final step (Fig.~\ref{segment_steps}.d) implements a contour-based erosion approach, refining granule boundaries derived from the optimised segmentation map. White granules/segments represent structures that were split into new segments from the initial segmentation, while bright lines within segments indicate regions that were merged into one segment compared to the initial segmentation\footnote{A preliminary code for the segmentation algorithm can be found at \url{https://github.com/Hypnus1803/SegmentPy}}.

For each segment, we calculated the area, eccentricity, and orientation. We also determined the mean intensity value of the blue-continuum, the mean strength of the magnetic field ($B_\mathrm{strength}$), the mean vertical magnetic field ($B_\mathrm{vertical}$), the mean horizontal magnetic field ($B_\mathrm{horizontal}$), the mean LOS velocity and the circular average of the magnetic field azimuth \citep[Eq. 2.2.4][]{Mardia1999}. Finally, we clean the dataset by removing outliers related to the area of the segments and any entries containing NaN values. The outliers were removed using standard score or z-score \citep{Kreyszig}, where values with $|zscore(\texttt{areas})|>3$ are considered outliers. To retain more data points from the other parameters, we set the outlier threshold to twice the z-score limit ($\sim 13.5$ arcsec$^2$).

\subsection{Eccentricity and orientation}
\label{sub_ecc}
Using standard libraries \citep[i.e. scikit-image][]{skimage2014}, we compute most of the morphological properties, including the eccentricity. The initial standard method used for eccentricity calculation depends on the length of the major and minor axes, which are derived from the second central moments of each region by fitting the best ellipse to each segment. Since solar granular patterns do not have perfect circular or elliptical shapes, this method for the calculation of the eccentricity could present certain issues. In addition, the fact that segmentation eventually fails to extract granules properly adds extra difficulties in these calculations.

Hence, we introduce two custom approaches to determine the eccentricity of granules. The first approach consists of three steps: first, we compute the geometrical centroid of the region or segment, defined as a point within the granule that is most distant from its boundary, calculated ``by replacing each foreground (non-zero) element, with its shortest distance to the background (any zero-valued element)''\footnote{definition taken from the library documentation in \url{https://docs.scipy.org/doc/scipy/reference/generated/scipy.ndimage.distance_transform_edt.html}} instead of using the geometrical shape itself to avoid centroids outside the segments; then, we calculate the line connecting the two most distant points that passes simultaneously through the centroid, namely the major axis. Finally, we calculate the line perpendicular to the major axis through the centroid, namely the minor axis. For the second approach, we first apply a convex hull transformation \citep[see Chapter 11;][]{Berg2010} to each segment and repeat the previous steps. In Appendix \ref{ecc} we describe and show the differences between each eccentricity calculation approach.

As the final data product, we compute the angular difference between the orientation of each granule's major axis obtained from the previous calculations and its circularly averaged magnetic field azimuth. Since we consider the relative difference between these variables, we calculate the angular difference within a range of $0^\circ$ (when the magnetic field azimuth and the major axis are parallel) to $90^\circ$ (when they are perpendicular).

The implementation and adaptation of all computational procedures, and data analysis were performed using custom Python scripts, incorporating also the specialized solar physics package Sunpy \citep{sunpyv5} for enhanced functionality.

\section{Results}

\begin{figure}[t]
\centering
   \includegraphics[width=9cm]{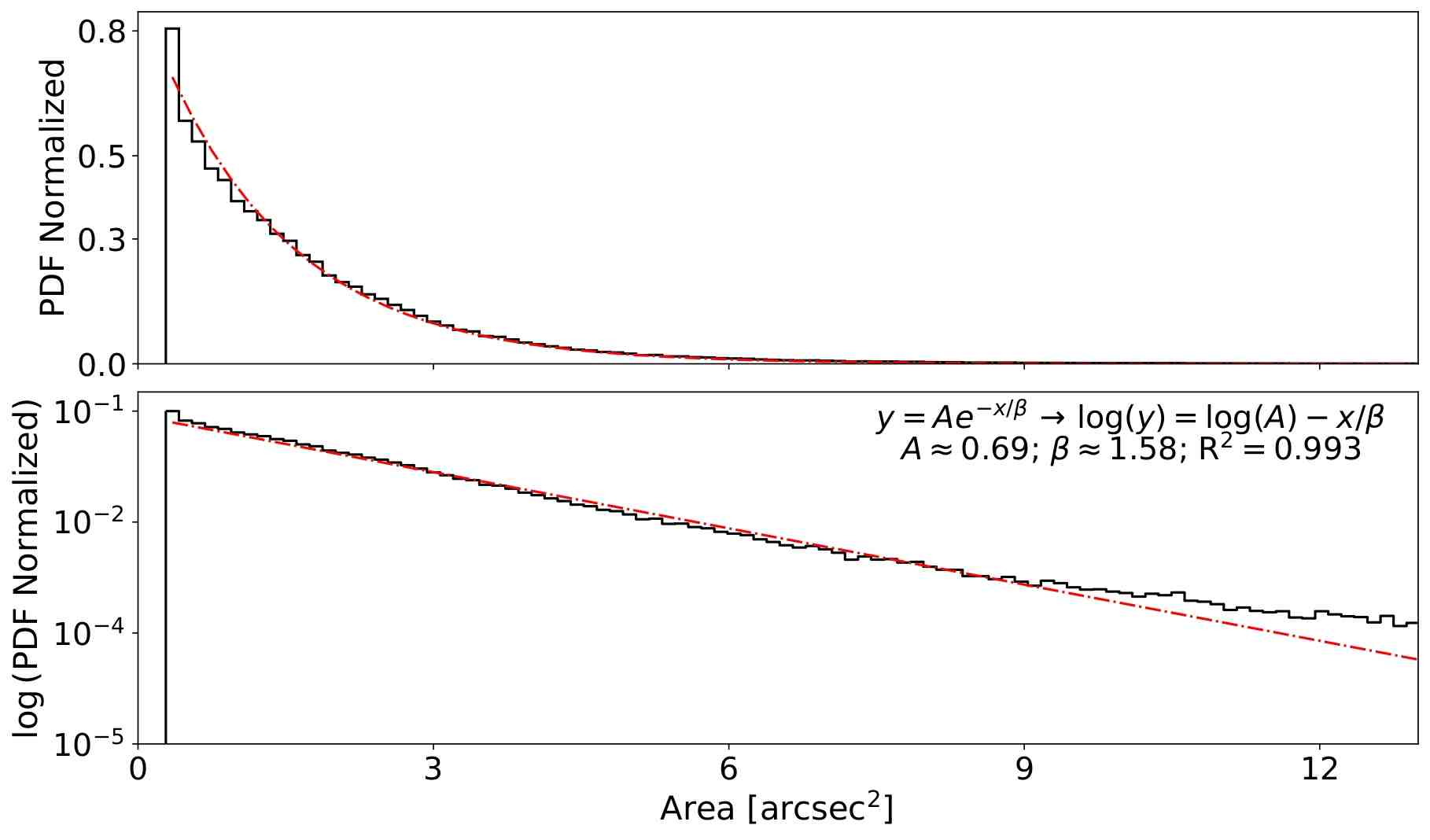}
      \caption{Granular area distributions. Top panel: raw distribution of granular areas including outliers, with a fixed bin width of $\sim0.15$ arcsec$^2$. Bottom panel: Histogram $\log(PDF)$ vs Area, where $PDF$ is the normalised probability density of the areas distribution. The dashed red line represents in each plot the fitting of the distribution with the found para\-meters.}
         \label{areas}
\end{figure}

Figure \ref{areas} (top panel) presents the area distribution of all identified granules. The temporal evolution of individual granules is not tracked; consequently, each granule may be counted multiple times throughout its lifespan. The resulting Probability Density Distribution (PDF) follows a negative exponential distribution where the lower limit of the granule sizes is set manually to $\sim 0.28$ arcsec$^2$, which also corresponds to the modal area of the whole dataset. To fit the PDF, we apply the natural logarithm to the PDF, and use a standard linear regression to recover the parameters that fit the exponential distribution. This exponential fit to our sample is valid for granules with sizes up to 10~arcsec$^2$, i.e., it fits $99.5 \%$ of the $725,044$ identified granules. The statistical sample is insufficient for larger granules, counting $3,453$ cases, thus we cannot claim that the power law is no longer valid for segmented areas larger than 10 arcsec$^2$. To fit the exponential distribution, we linearise it in the form

\begin{equation}
    PDF_{\text{area}} = A e^{\frac{-x}{\beta}} \, \xrightarrow{\text{$\log$}} \, \log PDF_{\text{area}} = \log(A) -\frac{x}{\beta},
    \label{eq1}
\end{equation}

where $PDF_{\text{area}}$ is the probability density distribution of the granule areas, $A$ is the amplitude or scaling factor which adjusts the height of the distribution ($A$ is expected to be $\sim1/\beta$), and the parameter $\beta$ (area units) is interpreted as the mean (expected value) of the distribution \citep[see ][for more details on the exponential distribution]{Balakrishnan96}. We found from fitting the right side of Eq.~\ref{eq1} (bottom panel in Fig.~\ref{areas}) that the mean area is $\sim 1.58$ arcsec$^2$ or an effective diameter of $\sim 1.42$ arcseconds which is close to $d=1.31$ arcseconds reported by \citet{Roudier1986}.

To better illustrate the importance of different eccentricity approaches described in Sect.~\ref{sub_ecc} and in Appendix \ref{ecc}, we present the relationship between granule areas and the eccentricities calculated using the three different methods through 2-dimensional kernel density estimation (2D KDE) plots. We use a grid size (number of bins) and number of levels of 100, which corresponds to a bin width of approximately half of the lower limit for the granule area mentioned previously ($\sim 0.14$ arcsec$^2$). To highlight regions of the KDE which have significantly lower population compared to the maximum of the KDE, we apply the logarithm to the distribution as illustrated in Fig.~\ref{areas}. Each KDE plot evaluates the joint distribution between the areas of the segments and the eccentricity measurements. This approach identifies patterns of the most common area and eccentricity values as well as their statistical relationship across the entire granule population.

\begin{figure*}[htb]
\centering
   \includegraphics[width=18cm]{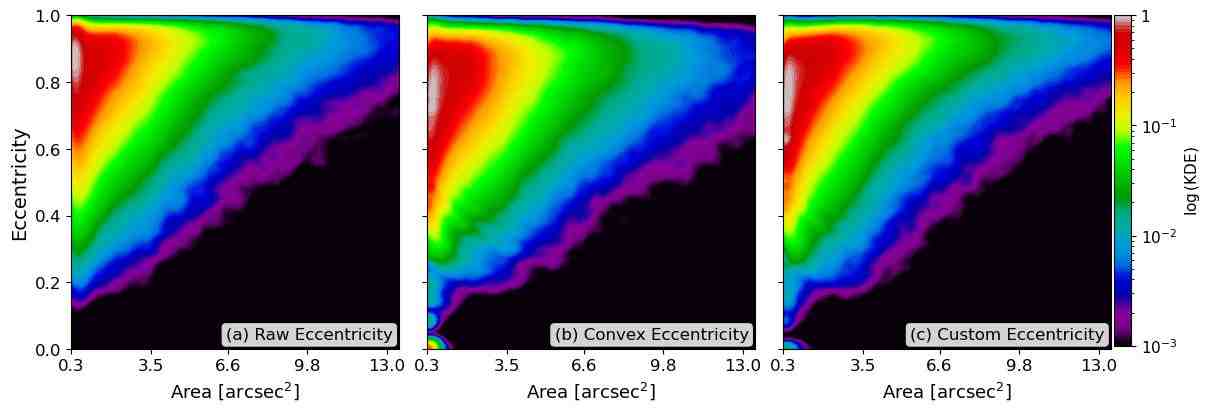}
      \caption{2D--$\log$(KDE) plots showing the distribution of the areas and three eccentricity metrics: (a) raw eccentricity (ellipse fitting), (b) convex eccentricity (computed on convex hulls), and (c) custom eccentricity (based on extreme points through the centre of mass of the segments). Each plot highlights the most probable eccentricity-area combinations and reveals the trends.}
         \label{kde_eccs}
\end{figure*}

Figure \ref{kde_eccs} displays the 2D KDE plots of the three different eccentricity approaches with the log-transformed kernel density, $\log(\mathrm{KDE})$. The left panel, using the raw eccentricity --defined by the best-fitting ellipse derived from the segment moments-- shows a high concentration of granules with eccentricities ranging between 0.84 and 0.91. This indicates that most granules appear highly elongated under this metric. Additionally, the plot shows a lower probability eccentricity boundary near 0.1 , with virtually no granules measured below this threshold.

The centre panel in Fig.~\ref{kde_eccs}, based on the convex-hull eccentricity, reveals density probabilities ranging eccentricities from 0.72 to 0.82.  This change toward lower eccentricity values implies this approach detects the presence of less elongated and more geometrically natural granule shapes, and also more ``circular'' granules appears because the plot shows density contours in lower eccentricity values around to zero. Although this behaviour may indicate a more responsive eccentricity measure, it is important to note that the convex hull transformation can sometimes introduce distortions in the granule shape representation (see Test segment 4 in Fig.~\ref{different_eccentricities}).

The third KDE plot (Fig.~\ref{kde_eccs}, right panel), using the custom eccentricity, presents a similar behaviour of the convex hull approach with a realistic representation. The highest density region is observed ranging values also from 0.72 to 0.83, but including also a local maximum around $\sim 0.63$, which is higher than the convex eccentricity but more sensitive to moderately elongated granule structures. This method also allows for eccentricity values approaching zero (circular shapes), offering a broader and more continuous range of morphological representations. Furthermore, the trend between areas and eccentricities shows that larger granules tend to exhibit higher eccentricities, since the most likely eccentricity for granules with areas greater than 3.5~arcsec$^2$ is around 0.9 that is larger than for smaller granules.

The colour-code of the KDE is normalised and reflects the statistical spread of the data, where black represents a probability of $10^{-3}$ or more dispersed data, while light-gray represents a probability of $1$, indicating high concentration in the combined data. Hereafter, the term \lq Eccentricity\rq\ is used for the custom eccentricity. Appendix \ref{ecc} describes in detail why we choose this approach.

\begin{figure*}[htb]
\centering
   \includegraphics[width=18.3cm]{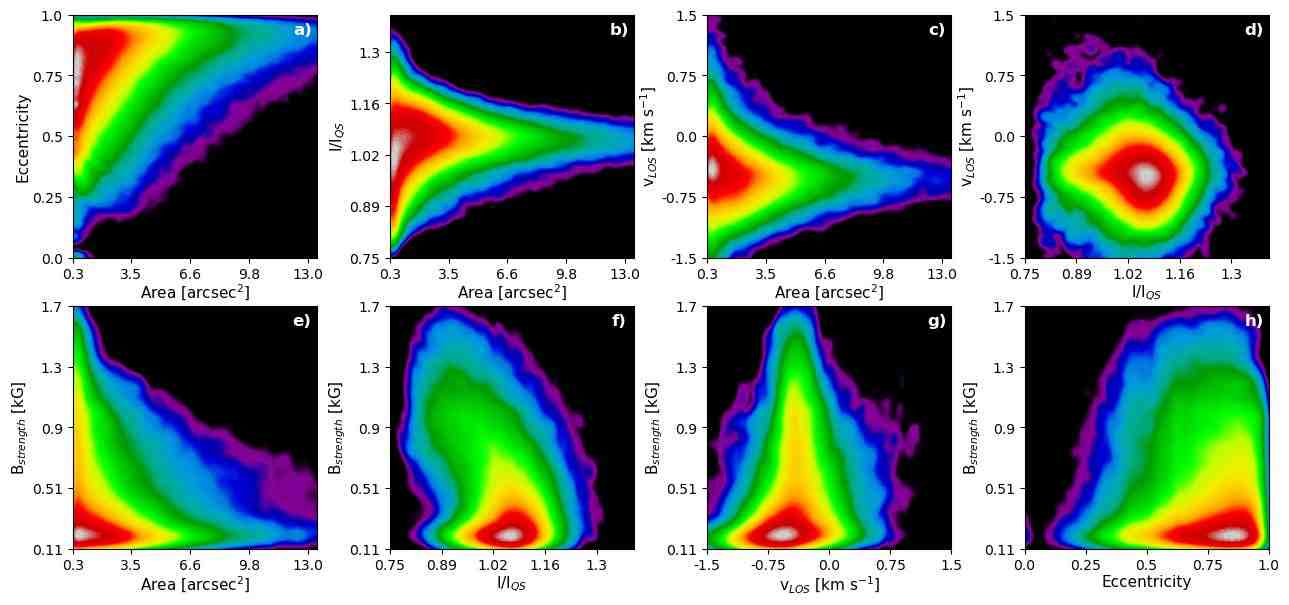}
      \caption{2D--$\log$(KDE) plots depicting the dependence of: (a) granular area versus granule eccentricity, (b) granular area versus granule continuum intensity, (c) granular area versus granule LOS velocities, and (d) granular continuum intensity versus granules LOS velocities. Bottom row shows the dependence of granule (e) area, (f) continuum intensity, (g) LOS velocity, and (h) eccentricity on magnetic field strength. Colour-coding follows the same scheme as in Fig.~\ref{kde_eccs}.}
         \label{globa_final}
\end{figure*}

The 2D-KDE plots in Fig.~\ref{globa_final} provide insight into several relationships between granular morphological properties and physical conditions. Table~\ref{table:1} summarises the maximum probability values in each KDE plot. The colour-coding in Fig.~\ref{globa_final} follows the same convention as Fig.~\ref{kde_eccs}, allowing consistent interpretation across figures.

Panel (a) in Fig.~\ref{globa_final} replicates panel  (c) in Fig.~\ref{kde_eccs}. The highest density region occurs near an eccentricity of ${\sim}0.77$ and an area value of ${\sim}0.55$ arcsec$^2$. We also see that larger granules tend to be more elongated.

\begin{table}[tb]
\caption{Maximum probability values determined by each 2D KDE.}
\small
\centering
\renewcommand{\arraystretch}{1.5}
\begin{tabular}{|c|c|c|c|c|c|c|c|}
\hline
\multicolumn{2}{|c|}{(a)} & \multicolumn{2}{c|}{(b)} & \multicolumn{2}{c|}{(c)} & \multicolumn{2}{c|}{(d)} \\
\hline
area & $e$ & area & $I/I_{QS}$ & area & $v_{LOS}$ & $I/I_{QS}$ & $v_{LOS}$ \\
0.55 & 0.77 & 0.55 & 1.03 & 0.55 & -0.41 & 1.08 & -0.47 \\
\hline
\hline
\multicolumn{2}{|c|}{(e)} & \multicolumn{2}{c|}{(f)} & \multicolumn{2}{c|}{(g)} & \multicolumn{2}{c|}{(h)} \\
\hline
area & $B$ & $I/I_{QS}$ & $B$ & $v_{LOS}$ & $B$ & $e$ & $B$ \\
0.68 & 0.205  & 1.07 & 0.185  & -0.56 & 0.191 & 0.85 & 0.185 \\
\hline
\end{tabular}
\tablefoot{Area values are presented in arcsec$^2$. Eccentricity (e) is dimensionless, as well as $I/I_\mathrm{QS}$. The $v_\mathrm{LOS}$ is expressed in km s$^{-1}$. The magnetic field strength $B\rightarrow B_\mathrm{strength}$ is measured in kG.}            
\label{table:1} 
\end{table}

Panel (b) in Fig.~\ref{globa_final} shows the correlation between the granular area and the continuum intensity, revealing that the size of the granules correlates with brightness. The maximum likelihood of granular brightness is independent on the size of the granules, it is ${\sim}1.03~I/I_\mathrm{QS}$. However, with increasing size of the granules, we observe a smaller scatter of their mean brightness (we note that this trend is primarily caused by smaller number of large granules). We note that the population of small and darker granules can be identified in the PDF as an extended red region for very small granules.

In panel (c) of Fig.~\ref{globa_final}, we examine the relationship between granular areas and LOS velocities. The resulting KDE is comparable to panel (b). We find that the maximum likelihood of LOS velocity is around $-0.4~\mathrm{km\,s^{-1}}$ for granules of all sizes. As in panel (b), smaller granules have higher spread of mean LOS velocity, but again this trend is primarily caused by smaller number of large granules 

Panel (d) in Fig.~\ref{globa_final} presents the relationship between granule continuum intensity and LOS velocity revealing a symmetric structure centred at $I/I_\mathrm{QS} \approx 1.08$ and $v_\mathrm{LOS} \approx -0.5~\mathrm{km s^{-1}}$. This 2D-KDE illustrates that mean brightness of granules does not obviously correlate with their mean LOS velocity, i.e., stronger up-flows in granules do not result into brighter granules. This does not contradict the intuitive expectation that brightest parts of the granules should have the strongest up-flows \citep[as shown for penumbral filaments by][]{Tiwari2013}, as we have to keep in mind that we are averaging both the brightness and the LOS velocity over the whole area of the segmented granule. We conclude that this averaging is exactly the cause of the loss of dependence between these parameters. 

In panel (e) of Fig.~\ref{globa_final}, we show the 2D-KDE between granules area and magnetic field strength. The most frequent $B$ found in the field of view is around 205~G and can be partially attributed to the noise level in the data as discussed in Sect.~\ref{inversions}. We observe that granular area decreases with increasing magnetic field strength. The largest granules are found in non-magnetic regions, while higher magnetic strengths are associated with significantly smaller convective cells. This suggests that magnetic fields inhibit the lateral expansion of granules. We do not observe any signature of equipartitional magnetic field strength ($B_\mathrm{eq} \sim 600$~G in the photosphere) in the KDE neither in panel (e) nor in panels (f, g, h) discussed later.

Panel (f) in Fig.~\ref{globa_final} shows the relationship between continuum intensity and magnetic field strength. The peak density occurs around $I/I_\mathrm{QS} \sim 1.07$ and $B \sim 185$~G. The KDE naturally captures fluctuations around the dominant trend, while the ridge of maximum probability density traces the most representative physical behaviour. As magnetic field strength increases, the continuum intensity decreases, providing evidence that strong magnetic fields suppress convective energy transport within these regions. 

Panel (g) illustrates the relationship between magnetic field strength and LOS velocity. The distribution is centred on upflows ($v_\mathrm{LOS} \approx -0.56$ km/s). With increasing field strength the distribution of LOS velocities narrows, but this is partially caused by smaller number of granules with high values of $B$. The KDE also shows that granules in regions with a weak magnetic field have an asymmetric distribution with preference for stronger up-flows, but this asymmetry disappears for granules located in regions with a stronger magnetic field.

Panel (h) in Fig.~\ref{globa_final} illustrates the magnetic field strength influence on the granule eccentricity. This plot reveals that high eccentricities dominate across the entire range of magnetic field strengths, with the highest probability density consistently concentrated toward eccentricity value $\approx 0.85$. The general trend of this plot shows that lower magnetic field strength allows for more circular shapes of granules.

\begin{figure}[!t]
\centering
   \includegraphics[width=9cm]{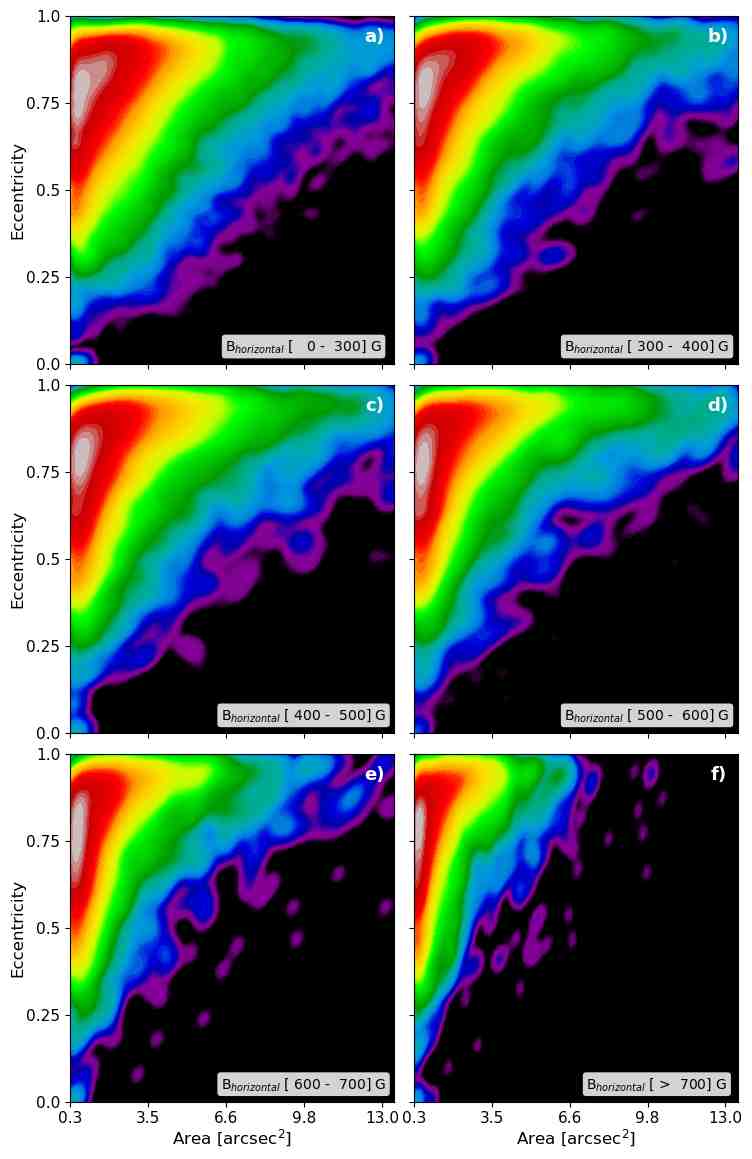}
      \caption{2D-KDE plots showing the dependence between area and eccentricity, segmented by horizontal magnetic field ($B_\mathrm{horizontal}$). Panel (a) ranges 0-300 G, representing  $\sim58\%$ of the detected granules. Panels (b-f) demonstrate how the area-eccentricity relationship changes across different horizontal magnetic field ranges. Colour-coding follows Fig.~\ref{kde_eccs}.}
         \label{area_vs_ecc}
\end{figure}

\begin{figure}[!t]
\centering
   \includegraphics[width=9cm]{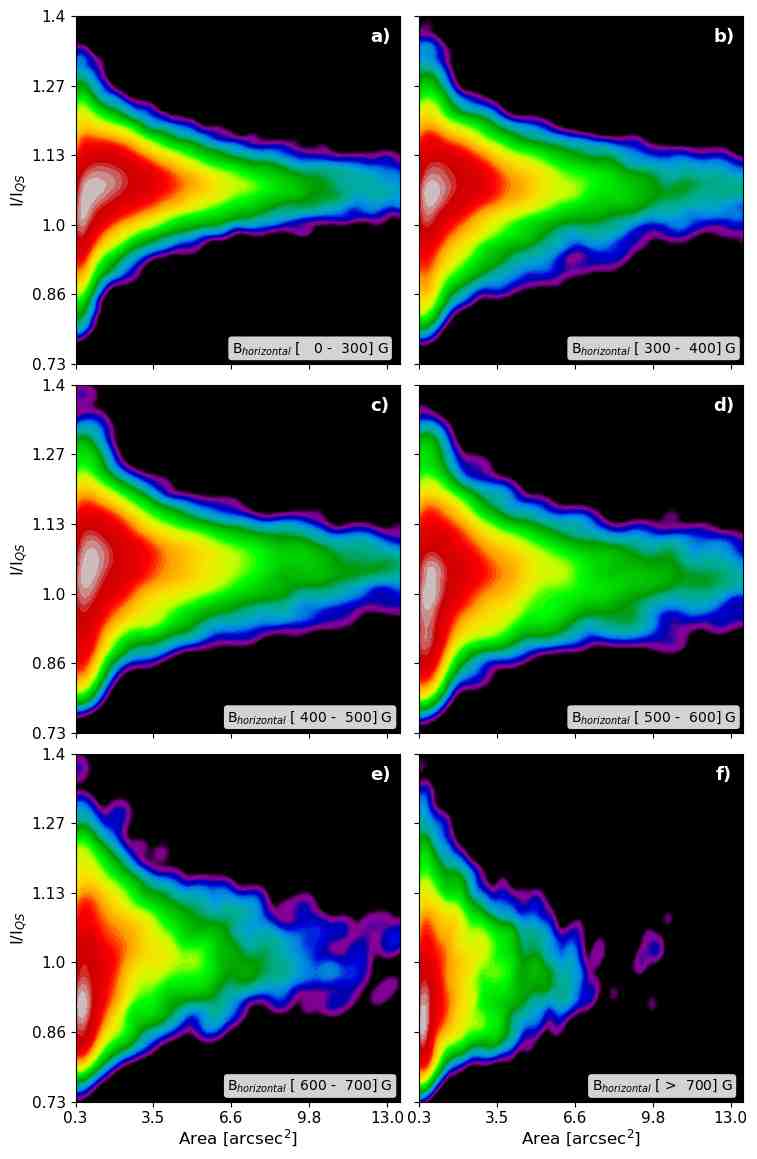}
      \caption{2D-KDE plots illustrating the relationship between granular area and continuum intensity and the dependence with the horizontal magnetic field. Panel (a) represents the 0-300 G range. Panels (b-f) show the evolution of the area-intensity relationship across other horizontal magnetic field ranges. Colour-coding follows Fig.~\ref{kde_eccs}.}
         \label{area_vs_i}
\end{figure}

The 2D-KDE plots in Figs.~\ref{area_vs_ecc},~\ref{area_vs_i},~\ref{angle_vs_area} show details of granular morphological properties and their relationship to different horizontal magnetic field regimes. Figure~\ref{area_vs_ecc} illustrates the dependence between granular areas and eccentricity as influenced by horizontal component of the magnetic field. Panel (a) corresponds to the 0–300 G range, which represents approximately 58\% of the data in the analysed FOV and closely resembles the global distribution shown previously in panel (c) of Fig.~\ref{kde_eccs}. This magnetic regime is typically associated with quiet-sun regions. As we advance through panels (b-f), we observe a progressive evolution in the area-eccentricity relationship. 

In regions of weak horizontal magnetic field, the distributions appear broad and the highest probability densities favour larger granules. As the horizontal magnetic field strength increases, the distributions shrink, with high-probability regions shifting toward smaller granular areas. This is in agreement with panel (e) of Fig.~\ref{globa_final}. Additional information provided in Fig.~\ref{area_vs_ecc} relates to the distribution of eccentricity based on the horizontal magnetic field strength. While the maximum likelihood of granules eccentricity does not depend on $B_\mathrm{hor}$, the granular areas begin to be truncated as $B_\mathrm{hor}$ increase, strongly shrinking the shape of the distribution. This shows that for strong magnetic fields, the density of large and elongated granules decreases whilst small granules are more likely to appear, a behaviour which is not evident from panel (h) of Fig.~\ref{globa_final}. 

Figure \ref{area_vs_i} explores the relation between the granular area and the continuum intensity across varying horizontal magnetic field strengths. Panel (a) corresponds to the 0-300 G range, and closely resembles the full distribution presented in Fig.~\ref{globa_final}~(b). Moreover, Fig.~\ref{area_vs_i} provides additional information compared to Fig.~\ref{globa_final}~(b) as it combines it with KDE shown in Fig.~\ref{globa_final}~(e) and~(f).

As the horizontal magnetic field increases through panels (b-f), systematic changes occur in the kernel density distributions. The analysis reveals that the ``tails'' of the 2D KDE distribution shift progressively downward with increasing horizontal magnetic field strengths, i.e., we do not observe large granules in regions of strong magnetic field as indicated by Fig.~\ref{globa_final} (e). We also see that the most common intensity of the granules decreases with increasing field strengths as indicated by Fig.~\ref{globa_final} (f). On the other hand, the probability of dark granules with $\sim 0.8~I/I_\mathrm{QS}$ is almost independent of the horizontal magnetic field, only in regions with $B_\mathrm{horizontal} > 600$~G it increases slightly. Also bright granules are observed in regions with strong magnetic field although their likelihood is smaller than in quiet Sun regions. This behaviour highlights the progressive suppression of both granule expansion and brightness as magnetic field strength increases, demonstrating the inhibitory effect of strong magnetic fields on convective processes within solar granulation.

Figure \ref{angle_vs_area} examines the relationship between granular eccentricity and the angular difference $\Delta\theta$ in different horizontal magnetic field regimes, where $\Delta\theta$ is explained in Sect.~\ref{sub_ecc}.
We note that values closer to $0^\circ$ suggest that the horizontal magnetic field affects the morphology parameters of the granules.

Panel (a) corresponds to regions with horizontal magnetic fields in the range of 0–300 G, characteristic of quiet-sun regions. The distribution appears uniform along $\Delta\theta$, indicating that granules of all eccentricities can be found across the entire range from parallel to perpendicular orientation in the differences, suggesting a weak coupling between granule orientation and azimuth of the magnetic field when the horizontal magnetic field is small. The same behaviour is found also for granules located in regions with $B_\mathrm{horizontal}$ between 300~G and 400~G. For stronger $B_\mathrm{horizontal}$, we find that the maximum likelihood of highly eccentric granules ($\approx 0.9$) is localized between $\Delta\theta = 10^\circ$ and $20^\circ$ suggesting that the magnetic field is aligned with the major axis of the granule in these magnetic regimes. For the strongest horizontal fields (panel e and f, $B_\mathrm{horizontal} > 600$~G, the alignment becomes less strong, likely reflecting the reduced number of granules observed within these intervals. We note that apart from the location and shape of the maximum likelihood (white-gray colour in the KDEs), the KDEs are not dependent on the $B_\mathrm{horizontal}.$

\section{Discussion and conclusions}

\begin{figure}[!t]
\centering
   \includegraphics[width=9.1cm]{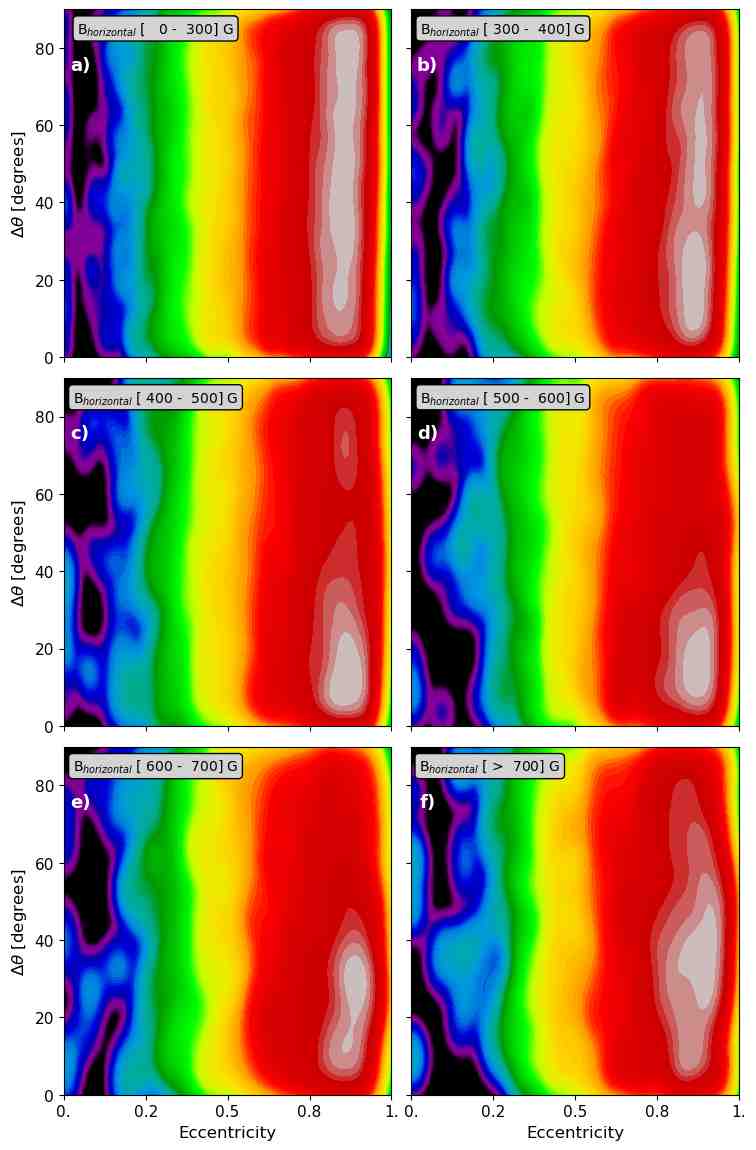}
      \caption{2D-KDE plots showing the relationship between granular area and $\Delta\theta$ as influenced by horizontal magnetic field. Panel (a) covers 0-300 G. Panels (b-f) illustrate how the area--$\Delta\theta$ relationship evolves across different horizontal magnetic field ranges. Colour-coding follows Fig.~\ref{kde_eccs}.}
         \label{angle_vs_area}
\end{figure}

A statistical study of convective cell properties in the presence of magnetic fields has been conducted. The analysis comprises the full range of granulation patterns observed in the SST field of view, from quiet Sun, through flux emergence regions, to a close neighbourhood of pores and sunspots. The findings corroborate and expand on previous case studies that examined convection under magnetic influence. For example, realistic 3D magneto-convection simulations demonstrate that strong magnetic fields concentrate into segmented structures and micropores along intergranular lanes, suppressing upflows, altering convective cell structure and dynamics from field-free cases \citep{Voegler2005}.

Regarding granular sizes, we measure a mean granule area of approximately 1.58 arcsec$^2$ with an effective diameter of 1.42 arcseconds, consistent with the diameter of 1.31 arcseconds reported by \citet{Roudier1986}. Concerning the relationship between granular morphology and magnetic field conditions, we confirm that granules are highly sensitive to the presence of the magnetic field. Our analysis demonstrates that granular area decreases systematically with increasing magnetic field strength. The largest granules are predominantly found in non-magnetic regions, while higher magnetic field strengths are associated with significantly smaller convective cells, indicating that relatively strong magnetic fields inhibit the lateral expansion of granules.

Both the mean continuum intensity of the granules and their size decrease systematically with increasing magnetic field strength as shown in Fig.~\ref{globa_final}~(e) and~(f) and in Fig.~\ref{area_vs_i}. This demonstrates the suppression of convective energy transport in magnetised regions, where stronger fields result in both smaller and dimmer granular structures.

Morphological analysis reveals that maximum granule eccentricity increases with increasing horizontal component of the magnetic field as shown in Fig.~\ref{area_vs_ecc}.  The probability of finding perpendicular alignment between the granule orientation and the magnetic field azimuth ($\Delta\theta$ approaching 90$^o$) decreases significantly as the horizontal field strength increases (Fig.~\ref{angle_vs_area}). This alignment effect is particularly prominent in flux emergence regions, where highly elongated granules connect opposite polarity regions. These morphological changes manifest the fundamental sensitivity of convective cells to magnetic field properties, as magnetic fields can significantly modify granule shapes and orientations.

Line-of-sight velocity analysis demonstrates that the majority of granules exhibit upflows, with the distribution centred around $-0.4~\mathrm{km\,s^{-1}}$. 
We also investigated in detail the role of the magnetic field on the LOS velocity. It turns out that the trend visible in Fig.~\ref{globa_final} (g) of smaller velocity span with increasing $|B|$ is caused by lower number of granules in regions with strong $|B|$ and we obtain comparable LOS velocities distributions for regions with different magnetic field strength (not displayed in any figure). We also want to stress that our statistical study shows that the mean brightness of granules does not correlate with their mean LOS velocity, i.e., stronger upflows in granules do not result into brighter granules. We want to stress that this finding does not contradict the intuitive expectation that brightest parts of granules should have the strongest up-flows as average values over the whole granule area are compared.

These findings demonstrate the mutual interaction between convective cells and magnetic fields. While magnetic fields suppress granular convection and modify granule morphology, the convective motions can also influence magnetic field configuration through processes such as line-of-sight field advection and magnetic flux concentration between granular boundaries. 

Future work will incorporate temporal evolution by applying a tracking algorithm to the segmented granules to investigate the dependence of convective cell lifetime on magnetic field strength and to analyse the evolution of magnetic field properties throughout individual granule lifespans. Such an approach will extend the present statistical analysis to a Lagrangian description of magneto-convection and allow us to directly link morphological changes to the temporal evolution of the magnetic field \citep[e.g. ][]{Zhang2009}. In addition, the inclusion of apparent horizontal granular motions and their dependence on magnetic field  will be particularly relevant. Previous studies indicate that such flows are important for plasma–magnetic field coupling and may have implications for energy transport into the upper solar atmosphere \citep[see ][]{Welsch2015,Tilipman2023}, while stronger magnetic fields tend to suppress horizontal motions \citep[e.g. ][]{Title1992,Aparna2025}. Examining how granular flows and morphology vary across different magnetic environments would help to further constrain the coupling between convection and magnetic fields in the quiet Sun.

\begin{acknowledgements}
This research is supported by the Czech--German common grant, funded by the Czech Science Foundation under the project 23-07633K and by the Deutsche Forschungsgemeinschaft under the project BE 5771/3-1 (eBer-23 13412) and the institutional support RVO:67985815. This research has made use of NASA's Astrophysics Data System Bibliographic Services. The research was sponsored by the DynaSun project and has thus received funding under the Horizon Europe programme of the European Union under grant agreement (no. 101131534). Views and opinions expressed are however those of the author(s) only and do not necessarily reflect those of the European Union and therefore the European Union cannot be held responsible for them.
\end{acknowledgements}

\bibliographystyle{aa} 
\bibliography{references} 

\begin{appendix}
\nolinenumbers
\section{Eccentricity}
\label{ecc}
\begin{figure}[H]
\centering
   \includegraphics[width=\linewidth]{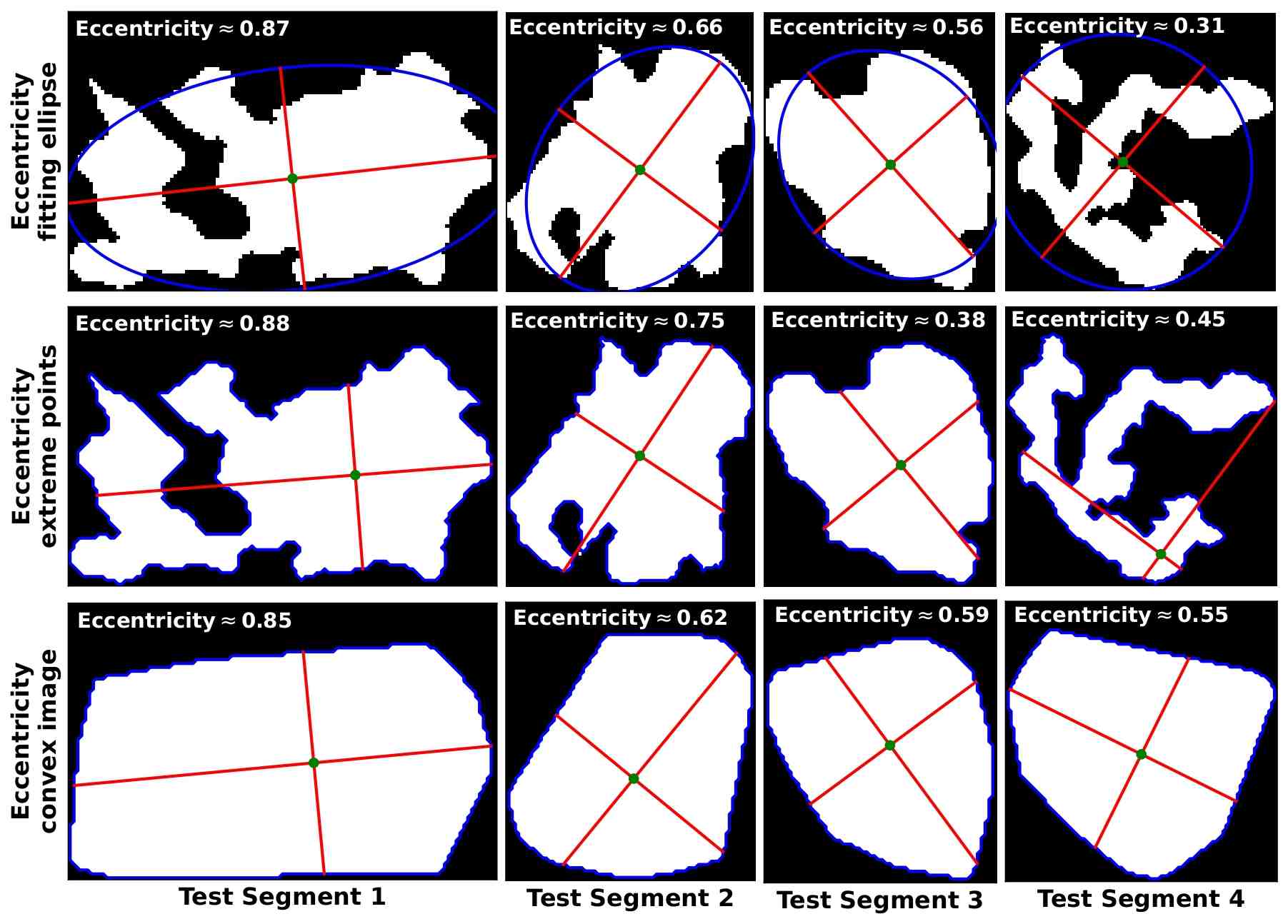}
      \caption{Comparison of three different eccentricity computation methods applied to four test segments. Top row: best-fit ellipse (image moments) approach; middle row: extreme points through centroid method; bottom row: extreme points through centroid method applied to convex hull image. The test segment 4 (rightmost column) corresponds to a special test showing a case where segmentation fails.
      }
         \label{different_eccentricities}
\end{figure}

The solar granules are changing constantly their shape. To study how the shape is changing when they are in presence/absence of magnetic field, it is necessary to include multiple morphological descriptors such as the area, the perimeter, the orientation, or elongation. The most used descriptor for the elongation measurement is the eccentricity, which can be defined on various ways \citep{Zhang2004,Rosin2005}, e.g. the factor between the minor and the major axis of the object or segment,
\begin{equation}
    e = \frac{\texttt{minor axis}}{\texttt{major axis}}.
\end{equation}
However, the most common definition of the eccentricity is such as the elongation of the ellipse defined as,
\begin{equation}
    e = \sqrt{1-\left(\frac{\texttt{minor axis}}{\texttt{major axis}}\right)^2}.
    \label{eq_e2}
\end{equation}
The main difference between both definitions is that Eq. \ref{eq_e2} distinguish better objects with larger elongations. In subsection \ref{sub_ecc}, we described three different methods calculate the eccentricities from the segmented granules. Figure \ref{different_eccentricities} shows examples of each eccentricity calculation approach. The columns depict four different isolated regions or segments labelled \lq Test Segment\rq\,1,2,3 and 4. Test Segment 4 shows an example where segmentation fails to extract the granule properly. The rows show the different approaches for calculating eccentricities. The first row shows the \lq best ellipse\rq\,fit calculated from the image moments, the second row shows the extreme points approach passing through the centroid, and the third row shows a similar approach, but applied over the region with the convex hull transformation. The eccentricity values can be seen for each segment and each eccentricity calculation approach. 

To decide which method is better, further the visual decision explained in Fig.\ref{kde_eccs}, we use other parameters, global shape descriptors as circularity ($4\pi A/P^2$), where $A$ is the area of the segment and $P$ is its perimeter, and the solidity ($A/A_{\texttt{convex\_hull}}$) where $A_{\texttt{convex\_hull}}$ is the area of the convex hull image of the segment. We also includes more sophisticated metrics, such as the Hausdorff distance \citep{Taha2015} and the chamfer distance transform \citep{Chetverikov1999}, which are more sensitive to shape irregularities. With all those parameters, we calculated the pearson and spearman correlation coefficients with the multiple shape descriptors summarised in Table \ref{table:A1}

\begin{table}[H]
\caption{Summarise of the correlations between each eccentricity $e$}
\centering
\renewcommand{\arraystretch}{1.5}
\begin{tabular}{|c|c|c|c|c|}
\hline
    & $p$ area & $s$ area & $p$ perimeter & $s$ perimeter\\
\hline
ellipse $e$ & 0.16 & 0.13 & 0.16 & 0.21\\
custom $e$ & 0.25 & 0.35 & 0.29 & 0.4\\
convex $e$ & 0.18 & 0.25 & 0.21 & 0.3\\
\hline
\hline
    & $p$ circularity & $s$ circularity & $p$ solidity & $s$ solidity\\
\hline
ellipse $e$ & -0.35 & -0.39 & -0.29 & -0.34\\
custom $e$ & -0.38 & -0.43 & -0.28 & -0.32\\
convex $e$ & -0.31 & -0.37 & -0.21 & -0.26\\
\hline
\hline
    & $p$ hausdorff & $s$ hausdorff & $p$ chamfer & $s$ chamfer\\
\hline
ellipse $e$ & 0.19 & 0.2 & 0.18 & 0.21\\
custom $e$ & 0.35 & 0.46 & 0.35 & 0.42\\
convex $e$ & 0.27 & 0.33 & 0.25 & 0.29\\
\hline
\end{tabular}
\tablefoot{ Correlation between each eccentricity approach and the different shape descriptors: area, perimeter, circularity, solidity, hausdorff distance and chamfer distance transform. Here $p$ and $s$ refers to the pearson and spearman correlation coefficients respectively.}            
\label{table:A1} 
\end{table}

The correlation analysis shows in Table \ref{table:A1} reveals that the custom eccentricity approach shows the best association with boundary irregularity measures, achieving spearman correlation of $0.458$ with hausdorff distance and $0.417$ with chamfer distance. These values indicate that custom eccentricity effectively ranks shapes according to their boundary deviations, which represents more realistic shapes of the granular segments. Additionally, the negative correlation with circularity (-0.434) confirms that the metric also captures deviations from circular geometry. However, the notable Spearman correlation with area (0.352) suggests a dependency on segment scale. This must be due to segmentation failure detecting large artifact granules.

The convex hull-based eccentricity approach exhibits moderate performance across all correlation measures, providing a balanced characterisation. While its association with the shape descriptors, circularity, solidity, hausdorff and chamfer distances, are less pronounced than those of the custom approach, this method demonstrates greater stability and reduced sensitivity to scale variations.

The ellipse-fitted eccentricity approach shows the weakest correlations with boundary irregularity measures, indicating limited capacity to detect structural deformations. Fitting an ellipse to the shape, smooths over boundary details and underestimates structural complexity. Consequently, while ellipse eccentricity may adequately describe simple elongation in smooth objects, it proves insufficient for applications for segments with irregular shape. 

This comparative analysis establishes custom eccentricity as the optimal choice for combined elongation and irregularity detection, with convex eccentricity serving as a suitable alternative when shape-elongation measurement without scale dependency is more important.

\end{appendix}
\end{document}